\documentclass[conference]{IEEEtran}
\IEEEoverridecommandlockouts
\usepackage{cite}
\usepackage{amsmath,amssymb,amsfonts}
\usepackage{algorithmic}
\usepackage{multirow}
\usepackage{graphicx}
\usepackage{textcomp}
\usepackage{xcolor}
\def\BibTeX{{\rm B\kern-.05em{\sc i\kern-.025em b}\kern-.08em
    T\kern-.1667em\lower.7ex\hbox{E}\kern-.125emX}}
\begin{document}

%\title{Autoencoder-Based DNNs for\\Ultrasound-Based Silent Speech Interfaces\\
\title{Autoencoder-Based Articulatory-to-Acoustic\\Mapping for Ultrasound Silent Speech Interfaces
\\
%{\footnotesize \textsuperscript{*}Note: Sub-titles are not captured in Xplore and
%should not be used}
%\thanks{László Tóth was supported by the János Bolyai Research Scholarship
%of the Hungarian Academy of Sciences and the UNKP-18-4 New
%Excellence Program of the Hungarian Ministry of Human Capacities.
%Tam\'as Gr\'osz was supported by the National Research, Development
%and Innovation Office of Hungary through the Artificial Intelligence
%National Excellence Program (grant no.: 2018-1.2.1-NKP-2018-00008).
%We acknowledge the support of the Ministry of Human Capacities,
%Hungary grant 20391-3/2018/FEKUSTRAT. The Titan X graphics card used
%in this research was donated by the Nvidia Corporation.}
}

\author{\IEEEauthorblockN{G\'abor Gosztolya}
\IEEEauthorblockA{\textit{MTA-SZTE Research Group} \\
\textit{on Artificial Intelligence}\\
Szeged, Hungary \\
ggabor@inf.u-szeged.hu} \and \IEEEauthorblockN{\'Ad\'am Pint\'er}
\IEEEauthorblockA{\textit{Institute of Informatics} \\
\textit{University of Szeged}\\
Szeged, Hungary \\
} \and \IEEEauthorblockN{L\'aszl\'o T\'oth}
\IEEEauthorblockA{\textit{Institute of Informatics} \\
\textit{University of Szeged}\\
Szeged, Hungary \\
tothl@inf.u-szeged.hu} \and \IEEEauthorblockN{Tam\'as Gr\'osz}
\IEEEauthorblockA{\textit{Institute of Informatics} \\
\textit{University of Szeged}\\
Szeged, Hungary \\
groszt@inf.u-szeged.hu} \and

\IEEEauthorblockN{Alexandra Mark\'o}
\IEEEauthorblockA{\textit{Department of Phonetics, } \\
\textit{E\"otv\"os Lor\'and University}\\
\textit{MTA-ELTE Lend\"ulet Lingual Articulation Research Group} \\
Budapest, Hungary \\
marko.alexandra@btk.elte.hu} \and \IEEEauthorblockN{Tam\'as G\'abor
Csap\'o}
\IEEEauthorblockA{\textit{Department of Telecommunications and Media Informatics,} \\
\textit{Budapest University of Technology and Economics}\\
\textit{MTA-ELTE Lend\"ulet Lingual Articulation Research Group} \\
Budapest, Hungary \\
csapot@tmit.bme.hu}
}

\maketitle

\begin{abstract}
When using ultrasound video as input, Deep Neural Network-based
Silent Speech Interfaces usually rely on the whole image to estimate
the spectral parameters required for the speech synthesis step.
Although this approach is quite straightforward, and it permits the
synthesis of understandable speech, it has several disadvantages as
well. Besides the inability to capture the relations between close
regions (i.e. pixels) of the image, this pixel-by-pixel
representation of the image is also quite uneconomical. It is easy
to see that a significant part of the image is irrelevant for the
spectral parameter estimation task as the information stored by the
neighbouring pixels is redundant, and the neural network is quite
large due to the large number of input features. To resolve these
issues, in this study we train an autoencoder neural network on the
ultrasound image; the estimation of the spectral speech parameters
is done by a second DNN, using the activations of the bottleneck
layer of the autoencoder network as features. In our experiments,
the proposed method proved to be more efficient than the standard
approach: the measured normalized mean squared error scores were
lower, while the correlation values were higher in each case. Based
on the result of a listening test, the synthesized utterances also
sounded more natural to native speakers. A further advantage of our
proposed approach is that, due to the (relatively) small size of the
bottleneck layer, we can utilize several consecutive ultrasound
images during estimation without a significant increase in the
network size, while significantly increasing the accuracy of
parameter estimation.
\end{abstract}


\section{Introduction}

%During the last several years, there has been significant interest
%in the articulatory-to-acoustic conversion research field, which is
%often referred to as ``Silent Speech Interfaces''
%(SSI)~\cite{Denby2010}. This has the main idea of recording the
%soundless articulatory movement, and automatically generating speech
%from the movement information, while the subject is not producing
%any sound. The reverse problem is called 'acoustic-to-articulatory
%inversion'. In the recent years, deep learning based methods have
%appeared in both fields.
%
%\subsection{Articulatory-to-Acoustic Mapping}

Over the last decade, there has been an increased interest in the
analysis, recognition and synthesis of silent speech, which is a
form of spoken communication where an acoustic signal is not
produced; that is, the subject is just silently articulating without
producing any sound. Systems which can perform the automatic
articulatory-to-acoustic mapping are often referred to as ``Silent
Speech Interfaces" (SSI)~\cite{Denby2010}. Such an SSI can be
applied to help the communication of the speaking impaired (e.g.
patients after laryngectomy), and in situations where the speech
signal itself cannot be recorded (e.g. extremely noisy environments
or certain military applications).

In the area of articulatory-to-acoustic mapping, several different
types of articulatory tracking equipment types have already been
used, including ultrasound tongue imaging
(UTI)~\cite{Denby2004,Denby2011,Hueber2010,Hueber2011,Jaumard-Hakoun2016,csapo2017dnnbased,grosz2018f0estimation,toth2018multitask,Xu2017,Tatulli2017},
electromagnetic articulography
(EMA)~\cite{Wang2012a,Wang2014,Bocquelet2016,Kim2017a,Cao2018,Taguchi2018},
permanent magnetic articulography
(PMA)~\cite{Fagan2008,Gonzalez2017a}, surface electromyography
(sEMG)~\cite{Nakamura2011,Deng2014,Diener2015,Janke2017,Meltzner2017,Diener2018a,Wand2018},
and Non-Audible Murmur (NAM)~\cite{Shah2018}. Of course, the
multimodal combination of these methods is also
possible~\cite{Freitas2014}, and the above methods may also be
combined with a simple video recording of the lip
movements~\cite{Hueber2010,Wand2016}.

There are basically two distinct ways of SSI solutions, namely
`direct synthesis' and
`recognition-and-synthesis'~\cite{Schultz2017a}. In the first case,
the speech signal is generated without an intermediate step,
directly from the articulatory data, typically using vocoders~
\cite{Denby2004,Hueber2011,Jaumard-Hakoun2016,csapo2017dnnbased,grosz2018f0estimation,Bocquelet2016,Gonzalez2017a,Diener2015,Janke2017,Cao2018,Taguchi2018,Diener2018a}.
In the second case, silent speech recognition (SSR) is applied on
the biosignal which extracts the content spoken by the person (i.e.,
the result is text). This step is then followed by text-to-speech
(TTS)
synthesis~\cite{Hueber2010,Denby2011,Tatulli2017,Wang2012a,Wang2014,Kim2017a,Fagan2008,Meltzner2017,Wand2018}.
A drawback of the SSR+TTS approach might be that the errors made by
the SSR component inevitably appear as errors in the final TTS
output~\cite{Schultz2017a}, and also that it causes a significant
end-to-end delay. Another drawback is that any information related
to speech prosody is totally lost, while several studies have showed
that certain prosodic components may be estimated reasonably well
from the articulatory recordings (e.g.,
energy~\cite{csapo2017dnnbased} and
pitch~\cite{grosz2018f0estimation}). Also, the smaller delay got by
using the direct synthesis approach may enable conversational use
and allows potential research on human-in-the-loop scenarios.
Therefore, state-of-the-art SSI systems mostly prefer the `direct
synthesis' principle.

\subsection{Deep Neural Networks for Articulatory-to-Acoustic Mapping}

As deep neural networks (DNNs) have become dominant in more and more
areas of speech technology, such as speech
recognition~\cite{Hinton2012}, speech synthesis~\cite{Ling2015} and
language modeling~\cite{Arisoy2012}, it is natural that the recent
studies have attempted to solve the acoustic-to-articulatory
inversion and articulatory-to-acoustic conversion problems using
deep learning.

For the task of articulatory-to-acoustic mapping, Diener and his
colleagues studied sEMG speech synthesis in combination with a deep
neural network~\cite{Diener2015,Janke2017,Diener2018a}. In their
most recent study~\cite{Diener2018a}, a CNN was shown to outperform
the DNN, when utilized with multi-channel sEMG data.
Domain-adversarial training, being a variant of multi-task training
was found to be suitable for adaptation in sEMG-based recognition
\cite{Wand2018}. Jaumard-Hakoun and her colleagues used a multimodal
Deep AutoEncoder to synthesize sung vowels based on ultrasound
recordings and a video of the lips~\cite{Jaumard-Hakoun2016}.
Gonzalez and his colleagues compared GMM, DNN and
RNN~\cite{Gonzalez2017a} models for PMA-based direct synthesis. We
used DNNs to predict the spectral
parameters~\cite{csapo2017dnnbased} and
F0~\cite{grosz2018f0estimation} of a vocoder using UTI as
articulatory input. Next, we expected that multi-task learning of
acoustic model states vs. vocoder parameters are two closely related
tasks over the same ultrasound tongue image input, and we found that
the parallel learning of the two types of targets is indeed
beneficial for both tasks~\cite{toth2018multitask}. Liu et al.
compared DNN, RNN and LSTM neural networks for the prediction of the
V/U flag and voicing~\cite{Liu2016}, while Zhao et al. found that
LSTMs perform better than DNNs for articulatory-to-F0
prediction~\cite{Zhao2017}. Similarly, LSTMs and bi-directional
LSTMs were found to be better in EMA-to-speech direct
conversion~\cite{Cao2018,Taguchi2018}. Generative Adversarial
Networks, a new type of neural network~\cite{Goodfellow2014}, were
also applied in the direct speech synthesis scenario, with promising
initial results~\cite{Shah2018}.

%\cite{Wand2016}
%\cite{Wand2018}
%\cite{Diener2018a}
%\cite{Cao2018}
%\cite{Shah2018}
%\cite{Taguchi2018}

%In the inversion field, researchers started to deal with deep
%architectures to accomplish an effective articulatory inversion.
%Uria and his colleagues compared a deep neural network and a deep
%trajectory mixture density network, and obtained better inversion
%accuracies with the latter than smoothing the results of a
%DNN~\cite{uria2012deep}. Wu and his colleagues used the DNN-based
%acoustic-to-articulatory inversion framework for a real-time
%speech-driven talking avatar system~\cite{wu2015acoustic}. Liu et
%al.\ tried bidirectional LSTMs and a deep recurrent mixture density
%network for the AAI task~\cite{liu2015deep}, which take into account
%the articulatory movements as a time series, as compared to the
%static DNN approaches before.

\subsection{Ultrasound Tongue Imaging}

\begin{figure}
\centering
\includegraphics[trim=0.3cm 8.3cm 0.3cm 1.2cm, clip=true, width=0.51\textwidth]{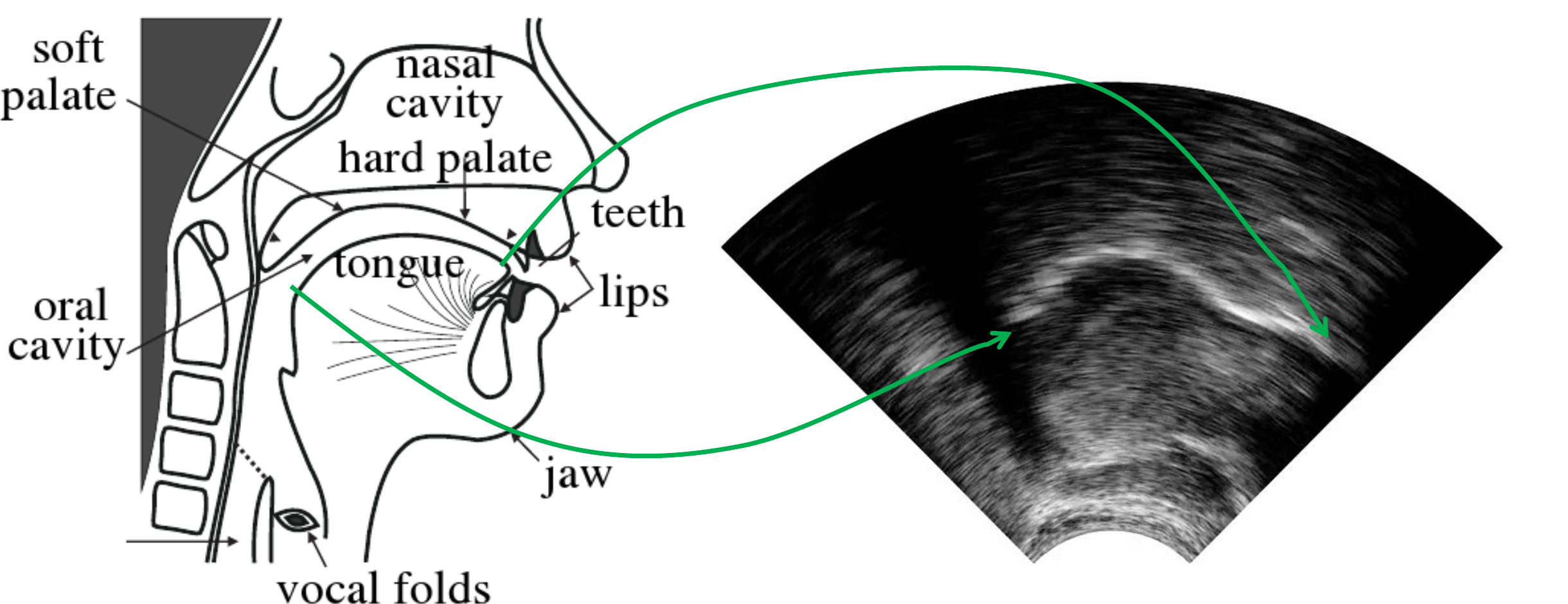}
\caption{\textit{Vocal tract (left) and a sample ultrasound tongue
image (right), with the same orientation.}} \label{fig:UTI_sample}
\end{figure}

Phonetic research has employed 2D ultrasound for a number of years
for investigating tongue movements during speech~\cite{Stone1983,
Stone2005a, Csapo2015a}. Usually, when the subject is speaking, the
ultrasound transducer is placed below the chin, resulting in
mid-sagittal images of the tongue movement. The typical result of 2D
ultrasound recordings is a series of gray-scale images in which the
tongue surface contour has a greater brightness than the surrounding
tissue and air. For a guide to tongue ultrasound imaging and
processing, see~\cite{Stone2005a}. A sample ultrasound image is
shown in Fig.~\ref{fig:UTI_sample}. UTI is a technique with higher
cost-benefit compared to other articulatory acquisition techniques,
if we take into account equipment cost, portability, safety and
visualized structures.

In the case of ultrasound-based SSI, the input of the machine
learning process is all the pixels of the ultrasound frame.
According to our earlier studies (see
e.g.~\cite{csapo2017dnnbased,grosz2018f0estimation,toth2018multitask}),
this approach is obvious and it allows the synthesis of intelligible
speech. However, it is suboptimal in many aspects. First, the input
image (in raw format $64\times946$, i.e. 60\,544 pixels) is highly
redundant, and contains a lot of irrelevant features -- which can be
partly managed by feature selection~\cite{csapo2017dnnbased}.
Second, the excessive number of features have a negative impact on
the effectiveness of the neural network (training and evaluation
time, number of stored weights), and they can also degrade the
predicted spectral parameters. With an efficient compression method,
both issues could be improved.

%Az ultrahangkép-alapú SSI esetében a gépi tanuló eljárás bemenetét
%egy képkocka pixelei jelentik. Könnyen látható, hogy ez a
%megközelítés, bár kézenfekvõ és korábbi tapasztalataink (ld.
%pl.~\cite{csapo2017dnnbased,grosz2018f0estimation,toth2018multitask})
%alapján érthetõ beszéd szintetizálását teszi lehetõvé, több
%tekintetben is szuboptimális. A bemenetként használt, képenként több
%ezer képpont (pl. a teljes nyers képkocka $64\times842$ méretû, azaz
%53\,888 képpontból áll) nagymértékben redundáns, valamint sok
%irreleváns jellemzõt is tartalmaz (bár ezen jellemzõkiválasztással
%lehet segíteni~\cite{csapo2017dnnbased}). A túl sok jellemzõ az
%alkalmazott mély háló hatékonyságára (tanítási és kiértékelési idõk,
%tárolt súlyok száma) egyértelmûen negatív hatással van, és a
%spektrális paraméterek becslését is ronthatja. Egy hatékony
%tömörítési eljárással mindkét területen javíthatunk.

\subsection{Current study}

In this study, we compress the input ultrasound images using an
autoencoder neural network. The estimation of the spectral speech
parameters is done by a second DNN, using the activations of the
bottleneck layer of the autoencoder network as features. According
to our experimental results, the proposed method is more efficient
than the standard approach, while the size of the DNN is also
significantly decreased.

%Jelen cikkünkben a bemenetként használt ultrahangképet egy
%autoenkóder hálózat segítségével tömörítjük, majd a beszédszintézis
%spektrális paramétereit a bottleneck réteg aktivációit mint
%jellemzõket használva becsüljük egy második mély neurális hálóval.
%Kísérleti eredményeink alapján a javasolt megközelítés pontosabb
%paraméterbecslést tesz lehetõvé, miközben a DNN mérete jelentõsen
%csökken.
\begin{figure*}[t]
\centering
\includegraphics[width=4cm]{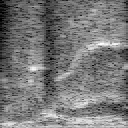}
\hspace{0.2cm}
\includegraphics[width=4cm]{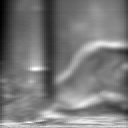}
\hspace{0.2cm}
\includegraphics[width=4cm]{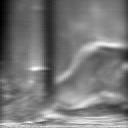}
\hspace{0.2cm}
\includegraphics[width=4cm]{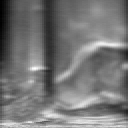}
\caption{A cavity ultrasound image in its original form (left), and
after encoding and restoring via an autoencoder network using
$N=64$, $N=256$ and $N=512$ neurons in the bottleneck layer.}
\label{pic_image_example}
\end{figure*}

\section{Estimating SSI Spectral Parameters Using Autoencoder Networks}

\subsection{Autoencoder}

Autoencoders (AE) are a special type of neural network that are used
to learn efficient data encodings in an unsupervised manner. They
are trained to restore the input values at the output layer; that
is, to learn a transformation similar to the identity mapping. This
forces the network to create a compact representation in the hidden
layer(s)~\cite{rumelhart1986learninginternal}. Technically, training
is usually realized by minimizing the mean squared error (MSE)
between its input and output, and the parameters can be optimized
via the standard back-propagation algorithm. Compression is enforced
by incorporating a {\it bottleneck layer}; i.e. a hidden layer,
which consists of significantly fewer neurons than the number of
input features (or the output layer). Previous studies have shown
that this technique can be applied to find relations among the input
features~\cite{lattner2017learningmusical}, for
denoising~\cite{geras2015scheduled},
compression~\cite{cheng2018deepconvolutional}, and even generating
new examples based on the existing
ones~\cite{zhao2017learninghierarchical}. Autoencoder neural
networks are used for example in image
processing~\cite{cheng2018deepconvolutional,chen2017stylebank},
audio processing~\cite{lattner2017learningmusical} and natural
language processing~\cite{andrews2016compressing}.

As for its structure, an autoencoder neural network consists of two
main distinct parts (cf. Fig.~\ref{dnn_layout}). The {\it encoder}
part is responsible for creating the compact representation of the
input, while the {\it decoder} part restores the input feature
values from the compact representation. The bottleneck layer is
located in the intersection of these two parts; the activations of
the neurons in this bottleneck layer can be interpreted as the
compact representation of the input. The encoder part can be viewed
as a dimension reduction method, which is trained together with the
reconstruction side (the decoder). After training, the encoder is
used as a feature extractor.

\subsection{Spectral Parameter Estimation by Autoencoder Neural Networks}
%\begin{figure}[t]
%\centering
%\includegraphics[width=12cm]{pics/dnn_layout_lncs_hun.eps}
%\caption{A javasolt kétlépéses DNN-alapú MGC-paraméterbecslõ eljárás
%mûködési sémája.} \label{dnn_layout}
%\end{figure}

\begin{figure}[!t]
\centerline{\includegraphics[width=8.5cm]{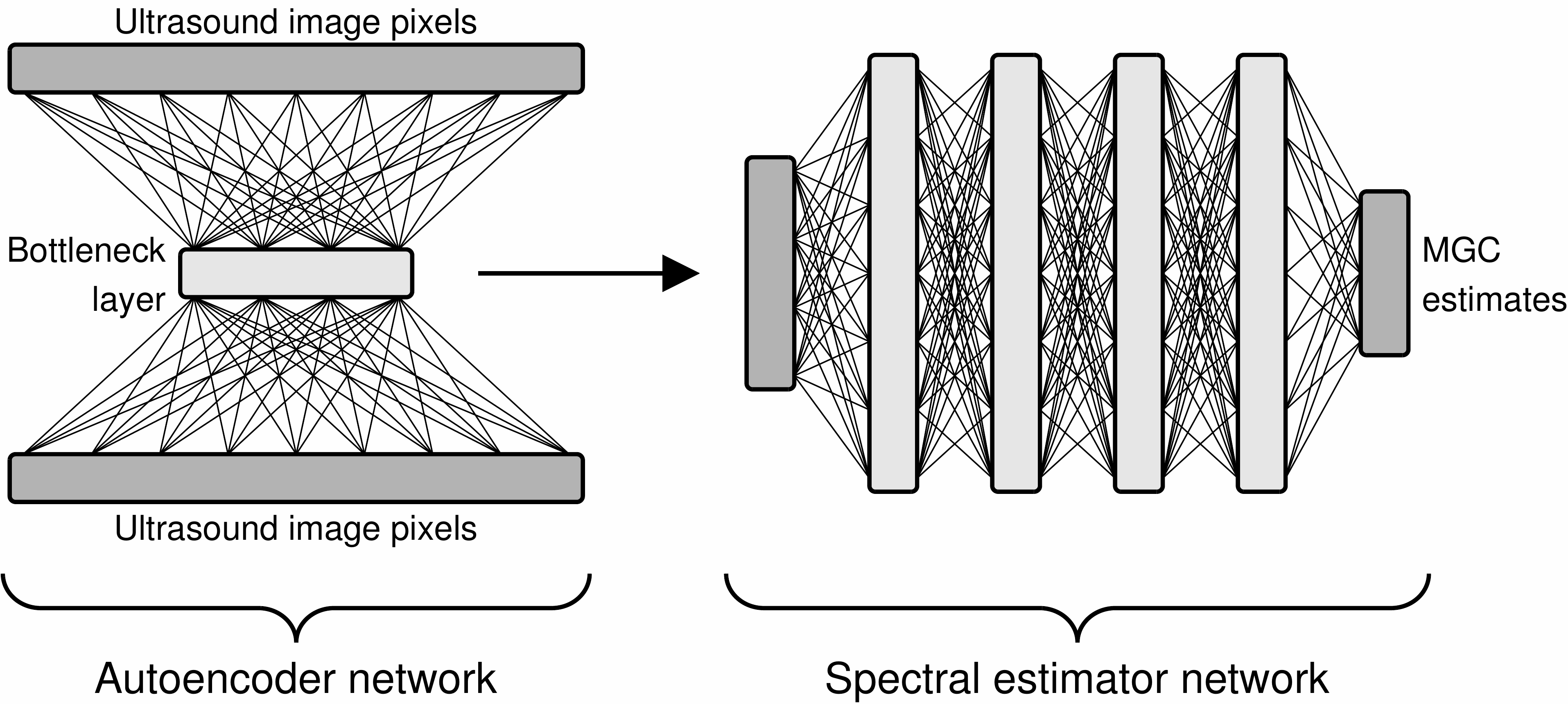}}
\caption{The workflow of the proposed DNN-based speech synthesis
(MGC-LSP) parameter estimation process.} \label{dnn_layout}
\end{figure}

In this study we propose to apply a two-step procedure to estimate
the speech synthesis spectral parameter values. In the first step we
train an autoencoder to reconstruct the pixel intensities of an
ultrasound image. Then, as the second step, we train another neural
network, this time just using the encoder part of the autoencoder
network to extract features. The task of this second network is to
learn the actual speech synthesis parameters (an {\it MGC-LSP}
vector and the gain) associated with the input ultrasound image.
(For the general scheme of the proposed method, see
Fig.~\ref{dnn_layout}.)
%\begin{figure*}[!t]
%\centerline{\includegraphics[width=12cm]{pics/dnn_layout.eps}}
%\caption{The workflow of the proposed DNN-based MGC-LSP parameter
%estimation process.} \label{dnn_layout}
%\end{figure*}

In our opinion, this approach has several advantages. One of them is
that the autoencoder network removes the redundancies present in the
image by finding the connections between different pixels of the
image. The second advantage is tied to the fact that the ultrasound
image is typically very noisy. Our expectation is that the
autoencoder network, by encoding only a limited amount of
information in its bottleneck layer, automatically performs some
kind of noise reduction, similar to the denoising autoencoder. A
third advantage of our approach might be that the bottleneck layer,
by nature, forces the network to compress the input images and keep
only the most important information. The usefulness of this
compression can be explained by the information bottleneck
theory~\cite{Tishby2015Deep}.

Using the output of the encoder part as features has another
practical advantage. Usually the number of weights in a standard
feed-forward DNN is significantly influenced by the number of input
features. For example, consider an input layer with 8\,192 neurons,
corresponding to the pixels of the ultrasound images (resized to
$64\times128$). Using 1\,024 neurons in the first hidden layer,
there will be roughly 8.4 million connections. Since the bottleneck
layer of the encoder contains considerably fewer neurons than the
first hidden layer of the estimator network, using it first to
extract features significantly reduces the size of our final
estimator network. This way the combined encoder and estimator
network becomes much smaller, which also speeds up inference. If we
follow the approach of our previous studies (see
e.g.~\cite{csapo2017dnnbased,grosz2018f0estimation,toth2018multitask}),
and also feed the feature vectors of the neighbouring images from
the video into the network, we can also apply a wider sliding window
without increasing the overall size of the network.

Fig.~\ref{pic_image_example} shows a sample ultrasound image in its
original form (left), and its reconstructions via three different
autoencoder networks, which differ only in the size of the
bottleneck layer (cases $N=64$, $N=256$ and $N=512$). It is quite
apparent that the original image is quite noisy, while the restored
images are much smoother. Furthermore, using more neurons in the
bottleneck layer preserves more image details. When we reduced the
size of the bottleneck layer, the restored image became blurrier,
and fine details were lost during the process. Of course, the
contour of the tongue is still quite distinct in all the images. It
is hard to determine, however, what level of detail is required for
optimal or close-to-optimal performance.

\section{Experimental Setup}

Next we describe the components of our experiments: the database we
used, the way we preprocessed the input image and the sound
recordings, and the meta-parameters of the neural network.

\subsection{Dataset}

The speech of one Hungarian female subject (42 years old) with
normal speaking abilities was recorded while she read 438 sentences
aloud. The tongue movement was also recorded in midsagittal
orientation using a ``Micro'' ultrasound system (Articulate
Instruments Ltd.) with a 2-4 MHz / 64 element 20mm radius convex
ultrasound transducer at 82 fps. During the recordings, the
transducer was fixed using an ultrasound stabilization headset
(Articulate Instruments Ltd.). The speech signal was captured with
an Audio-Technica - ATR 3350 omnidirectional condenser microphone
that was clipped approximately 20cm from the lips. Both the
microphone signal and the ultrasound synchronization signals were
digitized using an M-Audio -- MTRACK PLUS external sound card at
22\,050~Hz sampling frequency. The ultrasound and the audio signals
were synchronized using the frame synchronization output of the
equipment with the Articulate Assistant Advanced software
(Articulate Instruments Ltd.). The 438 recordings were split to form
a training set, a development set and a test set (310, 41 and 87
utterances, respectively).

\subsection{Preprocessing the speech signal}

For the analysis and synthesis of speech, a standard open source
vocoder was used from SPTK ({\tt http://sp-tk.sourceforge.net}).
%First, the speech signal was lowpass filtered and resampled at
%22\,050~Hz.
F0 was measured with the SWIPE algorithm~\cite{Camacho2008}. Next, a
24-order Mel-Generalized Cepstral analysis (MGC)~\cite{Tokuda1994}
was performed with $\alpha=0.42$ and $\gamma=-1/3$. MGCs were
converted to a Line Spectral Pair (LSP) representation, as these
have better interpolation properties. In order to synchronize the
result of the speech analysis with the ultrasound images, the frame
shift was chosen to be 1 / FPS (where FPS is the frame rate of the
ultrasound video). Together with the gain, the MGC-LSP analysis
resulted in a 25-dimensional feature vector, which was used in the
training experiments.

For the synthesis phase,
%we assumed that the F0 parameter can not be
%estimated from the articulatory images, so
we used the original F0 extracted from the input, which is standard
practice in standard SSI experiments (see
e.g.~\cite{Denby2004,Jaumard-Hakoun2016,csapo2017dnnbased,Diener2015}).
The predictions of the DNN served as the remaining MGC-LSP
parameters required by the synthesizer. First, impulse-noise
excitation was generated according to the F0 parameter. Afterwards,
spectral filtering was applied using the MGC-LSP coefficients and a
Mel-Generalized Log Spectral Approximation (MGLSA)
filter~\cite{imai1983mellog} to reconstruct the speech signal.

\subsection{Preprocessing the ultrasound signal}

The original ultrasound signal consisted of 64 beams, each having a
resolution of 946. First, we rearranged these signals to
64$\times$128 single-channeled images using a bicubic interpolation.
This reduction did not significantly affect the visual content of
the images, and the DNNs trained on these reduced images achieved
almost identical results~\cite{csapo2017dnnbased}. The original
pixels had an intensity in the range $[0, 255]$; following the
standard normalization technique in image processing (see
e.g.~\cite{varga2013informationcontent}), we divided the original
values by 255, converting them to the $[0, 1]$ scale in this way.

\subsection{DNN Parameters}

We implemented our neural networks in the Tensorflow
framework~\cite{tensorflow2015whitepaper}; the hidden layers
contained neurons using the Swish activation
function~\cite{ramachandran2018searching}, while the 25 output
neurons, corresponding to the speech synthesis spectral parameters,
were linear ones. We fixed the $\beta$ parameter of the Swish
neurons to $1.0$ (in this case, the Swish function is equivalent to
the sigmoid-weighted linear unit
(SiLU,~\cite{elfwing2018sigmoidweighted})). The loss function of the
network was the mean squared error, and it was minimized using the
Adam optimizer.

Our standard spectral estimator neural network, used as the
baseline, had input neurons which corresponded to the pixels of the
(resized) ultrasound video (8\,192 overall), while the five hidden
layers consisted of 1\,024 neurons each. We used L2 regularization
on the weights. From previous experience we know that incorporating
the features extracted from the neighbouring ultrasound images might
help in predicting the MGC-LSP
parameters~\cite{csapo2017dnnbased,grosz2018f0estimation,toth2018multitask};
hence we also trained a DNN which used the pixel values of five
consecutive images as its input (40\,960 input neurons in total).
The training targets were of course the MGC-LSP parameters
associated with the image located in the middle. These two DNNs had
12.6 million and 46.2 million weights overall, when using one and
five consecutive ultrasound images, respectively.

As regards the autoencoder network, we performed our experiments
using $N = 64$, $128$, $256$ and $512$ neurons in the bottleneck
layer; these were directly connected to the input and output layers,
without employing any further hidden layers. The input and output
layers of the autoencoder network corresponded to one ultrasound
image, so these contained 8\,192-8\,192 neurons. In the case where
the autoencoder bottleneck activations were used as input, the
spectral estimator DNN was a standard fully-connected feed-forward
DNN, having five hidden layers, each consisting of 1\,024 Swish
neurons. Notice that in this case the feature vector was an order of
magnitude smaller than that of the DNN trained on the original
network. This also allowed us to include several neighbouring
``images'' during DNN training and evaluation, so in this case, in
our experiments, we used a total of 1, 5, 9, 13 and 17 frames of the
ultrasound video during DNN training and evaluation.

\section{Results Using Objective Measurements}
\begin{figure*}[t]
\centering
\includegraphics[width=8.5cm]{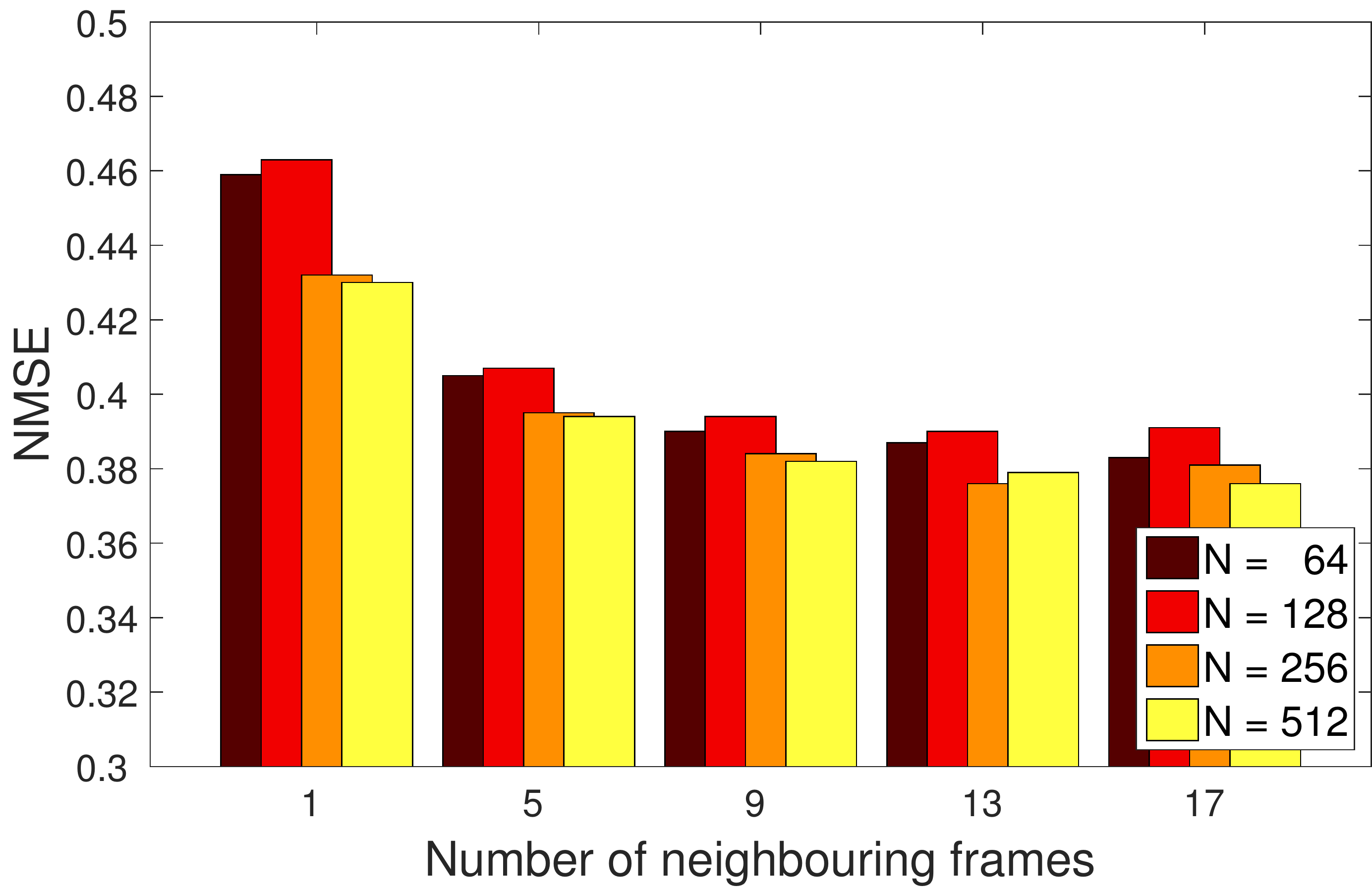}
\hspace{0.5cm}
\includegraphics[width=8.5cm]{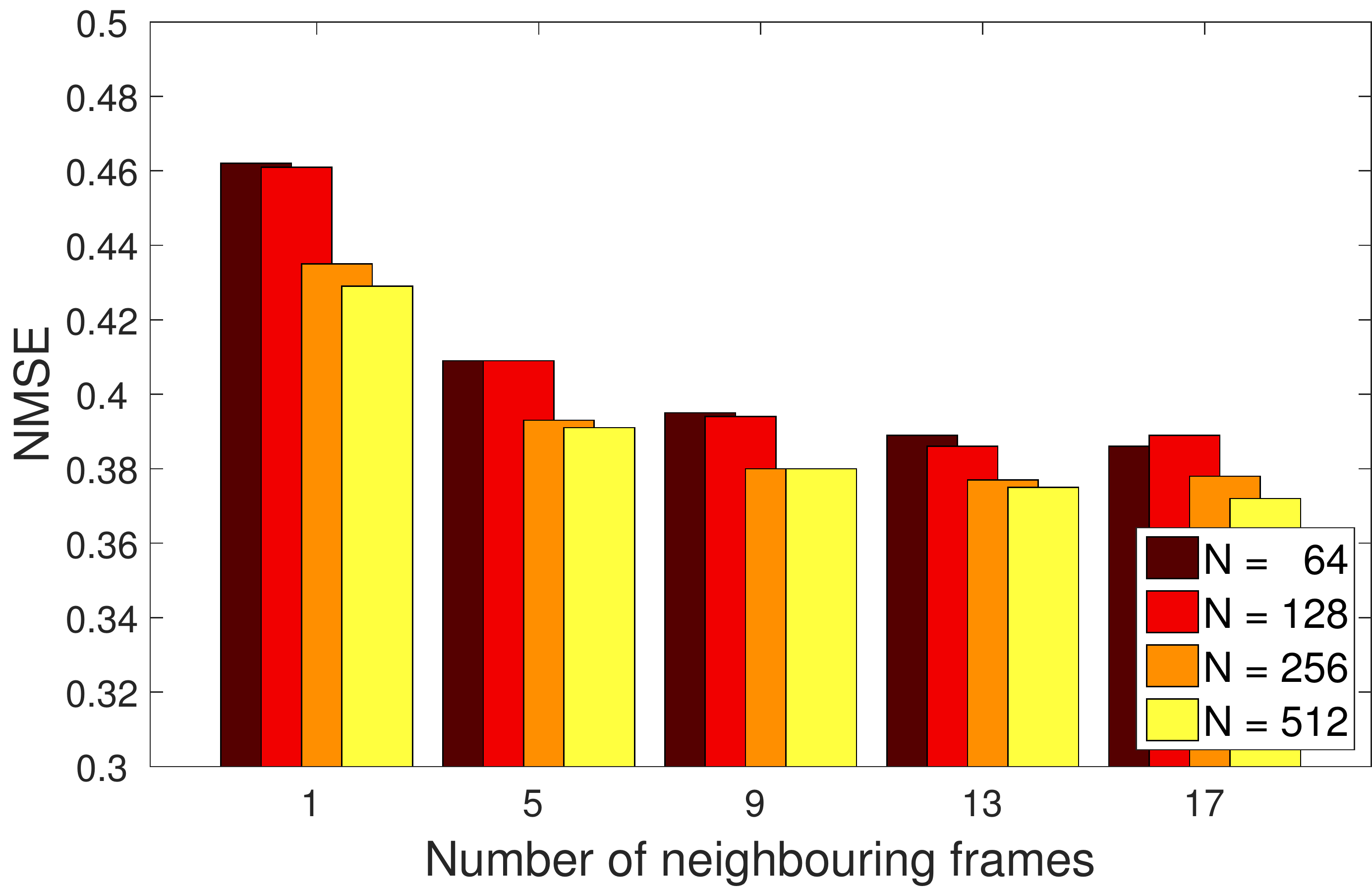}
\caption{Average NMSE scores measured on the development set (left)
and on the test set (right) as a function of the size of the
bottleneck layer of the autoencoder network ($N$) and the number of
neighbouring frames used.} \label{results_ae_nmse}
\end{figure*}
\begin{figure*}[t]
\centering
\includegraphics[width=8.5cm]{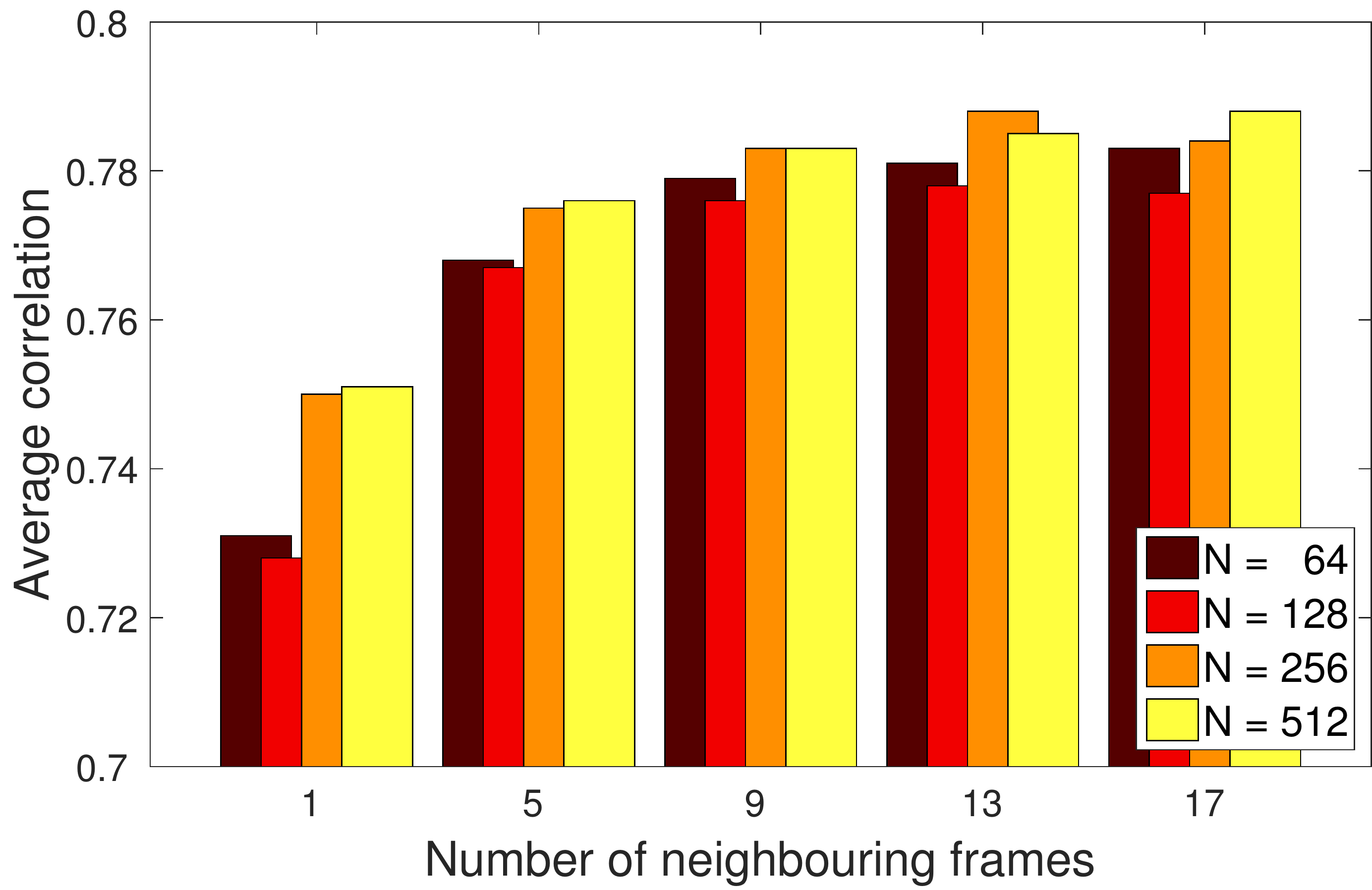}
\hspace{0.5cm}
\includegraphics[width=8.5cm]{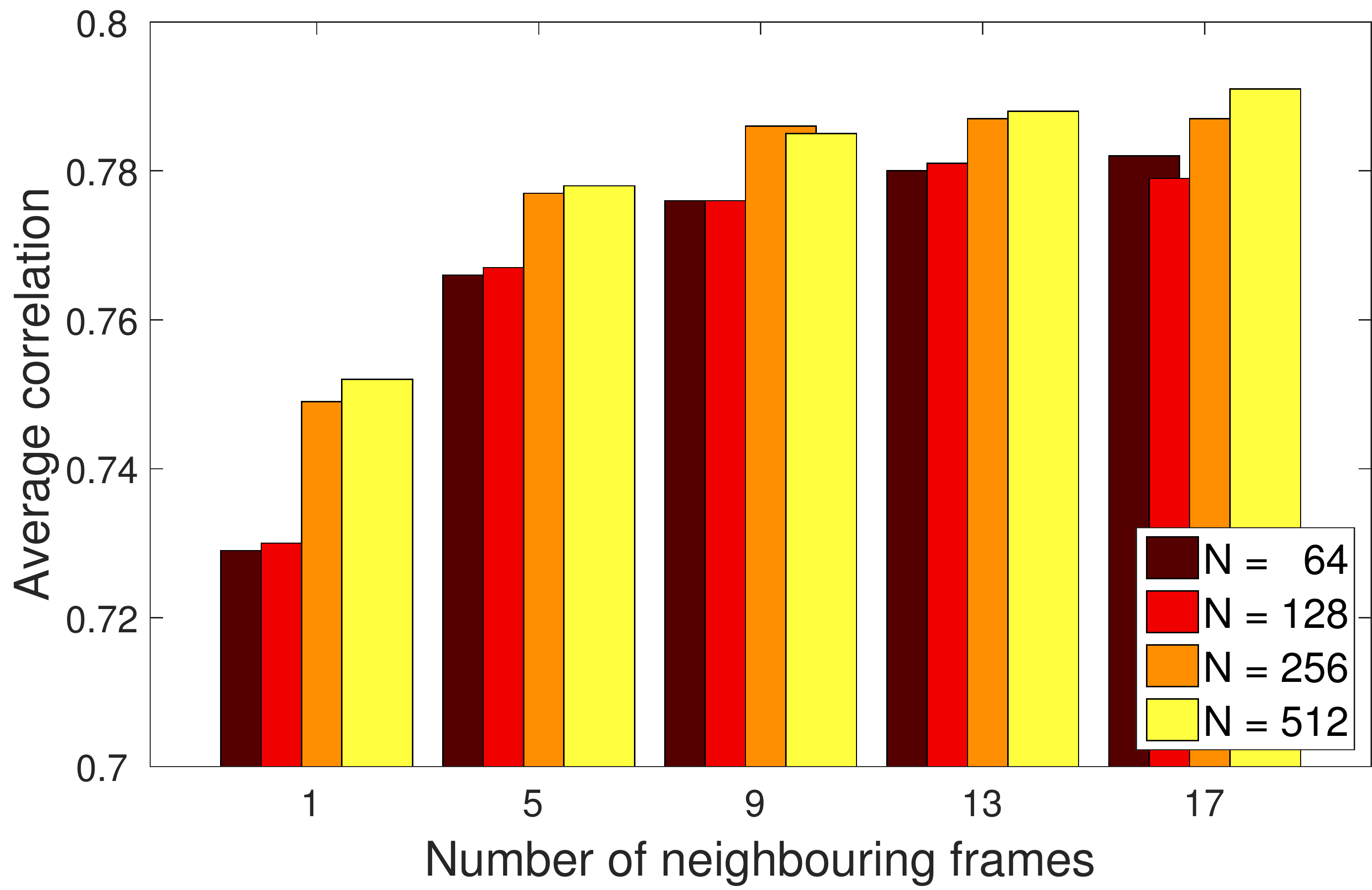}
\caption{Average Pearson's correlation scores measured on the
development set (left) and on the test set (right) as a function of
the size of the bottleneck layer of the autoencoder network ($N$)
and the number of neighbouring frames used.} \label{results_ae_corr}
\end{figure*}

Since estimating the MGC-LSP spectral parameters is a regression
task, first we evaluated the performance of the various models via
standard regression evaluation metrics. The first, quite
straightforward option is to use the Mean Squared Error (MSE); since
our DNN-based models predict 25 different speech synthesis
parameters, we took the average of the 25 MSE values. However, the
different output scores may have different ranges, which means that
a simple unweighted mean may be biased towards parameters operating
on a larger scale; to counter this effect, we used the Normalized
Mean Squared Error (NMSE) metric instead. Another evaluation metric
we applied was the Pearson's correlation of the original and the
estimated values; again, we simply averaged out the 25 correlation
scores obtained.
%\begin{figure}[t]
%\centering
%\includegraphics[width=8.5cm]{pics/results_nmse_test-eps-converted-to.pdf}
%\caption{Average NMSE scores measured on the test set as a function
%of the size of the bottleneck layer of the autoencoder network ($N$)
%and the number of neighbouring rows used.}
%\label{results_ae_nmse_test}
%\end{figure}
%\begin{figure}[t]
%\centering
%\includegraphics[width=8.5cm]{pics/results_corr_test-eps-converted-to.pdf}
%\caption{Average Pearson's correlation scores measured on the test
%set as a function of the size of the bottleneck layer of the
%autoencoder network ($N$) and the number of neighbouring rows used.}
%\label{results_ae_corr_test}
%\end{figure}

Fig.~\ref{results_ae_nmse} (left) shows the measured normalized mean
squared error scores on the development set for the different,
autoencoder network-based configurations. It is clear that, by using
1 and 5 (2-2) neighbouring frames, we get significantly worse
estimates than by using 9 (4-4) frames; when having a larger sliding
window size, however, the improvement becomes negligible. Examining
the size of the bottleneck layer of the autoencoder network we can
see that the networks having $N=64$ or $N=128$ neurons led to a
slightly less precise parameter estimates than with $N=256$ or
$N=512$; however, the difference was only significant when we did
not use any neighbouring frames. The NMSE scores measured on the
test set (see Fig.~\ref{results_ae_nmse} (right)) display
practically the same tendencies as those on the development set.

The mean Pearson's correlation scores behaved quite similarly both
on the development set (see Fig.~\ref{results_ae_corr} (left)) and
on the test set (Fig.~\ref{results_ae_corr} (right)): using 9 (4-4)
neighbouring feature vectors led to optimal or close-to-optimal
values. We found that it was worth employing at least 256 neurons in
the bottleneck layer of the autoencoder network, although the
observed difference was probably not significant among the different
configurations, at least when we relied on 9 or more neighbouring
images.

\begin{table*}[!t]
\caption{The average NMSE and average Pearson's correlation
coefficients measured on the development and test sets, and the
number of weights of the different configurations tested}
\begin{center}
\begin{tabular}{l|c|r||c|c||c|c}
                        &~No. of~       &~No. of~~  & \multicolumn{2}{c||}{~NMSE~} & \multicolumn{2}{c}{~Correlation~}\\
\cline{4-5}\cline{6-7}
Technique               &~frames~   &~weights~  &~Dev.~ &~Test~ &~Dev.~ &~Test~ \\
\hline\hline\multirow{2}{*}{Standard}
                        & ~1            &~12.6M~~   &~0.529~ &~0.534~ &~0.680~ &~0.676~ \\
                        & ~5            &~46.2M~~   &~0.523~ &~0.530~ &~0.684~ &~0.680~ \\
\hline\multirow{2}{*}{Autoencoder. N = ~64~}
                        & ~1            &~~4.8M~~   &~0.459~ &~0.462~ &~0.731~ &~0.729~ \\ % AE: 0.5M
                        & ~9            &~~5.3M~~   &~0.390~ &~0.395~ &~0.779~ &~0.776~ \\
\hline\multirow{3}{*}{Autoencoder. N = 256~}
                        & ~1            &~~6.6M~~   &~0.432~ &~0.435~ &~0.750~ &~0.749~ \\ % AE: 2.1M
%                        & ~5            &~~7.6M~~   &~0.395~ &~0.393~ &~0.775~ &~0.777~ \\
                        & ~9            &~~8.7M~~   &~0.384~ &~0.380~ &~0.783~ &~0.786~ \\
                        & 13            &~~9.7M~~   &~0.376~ &~0.377~ &~0.788~ &~0.787~ \\
\hline
                        & ~1            &~~8.9M~~   &~0.430~ &~0.429~ &~0.751~ &~0.752~ \\ % AE: 4.2M
Autoencoder. N = 512~   & ~5            &~11.0M~~   &~0.394~ &~0.391~ &~0.776~ &~0.778~ \\
                        & ~9            &~13.1M~~   &~0.382~ &~0.380~ &~0.783~ &~0.785~ \\
\end{tabular}
\label{table_res}
\end{center}
\end{table*}
Examining the actual normalized mean squared error and Pearson's
correlation values (see Table~\ref{table_res}) we notice that, when
we used the original ultrasound image pixel-by-pixel, the
neighbouring frame vectors did not help the prediction for some
reason (in our previous studies this was not the
case~\cite{csapo2017dnnbased,grosz2018f0estimation,toth2018multitask}).
Among the autoencoder-based models we achieved the best performance
for both objective evaluation metrics and for both subsets in the
$N=256$ case using 13 (6-6) neighbours; however, we also see that
using only 9 neighbouring frames leads to just slightly worse
scores. The NMSE scores of $0.376-0.394$ on the test set mean a
relative error reduction score of 25-29\%, while the $0.776-0.787$
correlation values brought relative improvements of 30-33\% over the
$0.680$ score used as the baseline; this improvement is definitely
significant.

Table~\ref{table_res} also lists the size (i.e. the total number of
weights) of each DNN model. Of course, for the autoencoder-based
models first we have to encode the ultrasound images; therefore, in
these cases, the indicated values already contain the size of the
encoding part of the autoencoder network (being 0.5 million
($N=64$), 1.0 million ($N=128$), 2.1 million ($N=256$) and 4.2
million ($N=512$)). It is quite apparent that the size of the
autoencoder-based models only rarely exceed the size of our baseline
model (which worked directly on the (resized) ultrasound image), and
they were significantly smaller in each case than the DNN working on
five consecutive frames. Based on these scores, we may conclude that
the proposed, autoencoder-based approach not only leads to a more
accurate estimation of the speech synthesis spectral parameters, but
it is also more feasible from a computational viewpoint.

\section{Subjective Listening Test Results}

In order to determine which proposed system is closer to natural
speech, we conducted an online MUSHRA (MUlti-Stimulus test with
Hidden Reference and Anchor) listening test~\cite{mushra}. The
advantage of MUSHRA is that it allows the evaluation of multiple
samples in a single trial without breaking the task into many
pairwise comparisons. %Our aim was simply to compare the natural
%sentences with the synthesized sentences with the baseline, the
%proposed approaches and a benchmark system (the latter having the
%same spectrum across the utterance).
In the test, the listeners had to rate the naturalness of each
stimulus in a randomized order relative to the reference (which was
the natural sentence), from 0 (very unnatural) to 100 (very
natural). We chose ten sentences from the test set; the variants
appeared in randomized order (different for each listener). Each
sentence was rated by 14 native Hungarian speakers.
\begin{figure}[t]
\centering
\includegraphics[width=8.5cm]{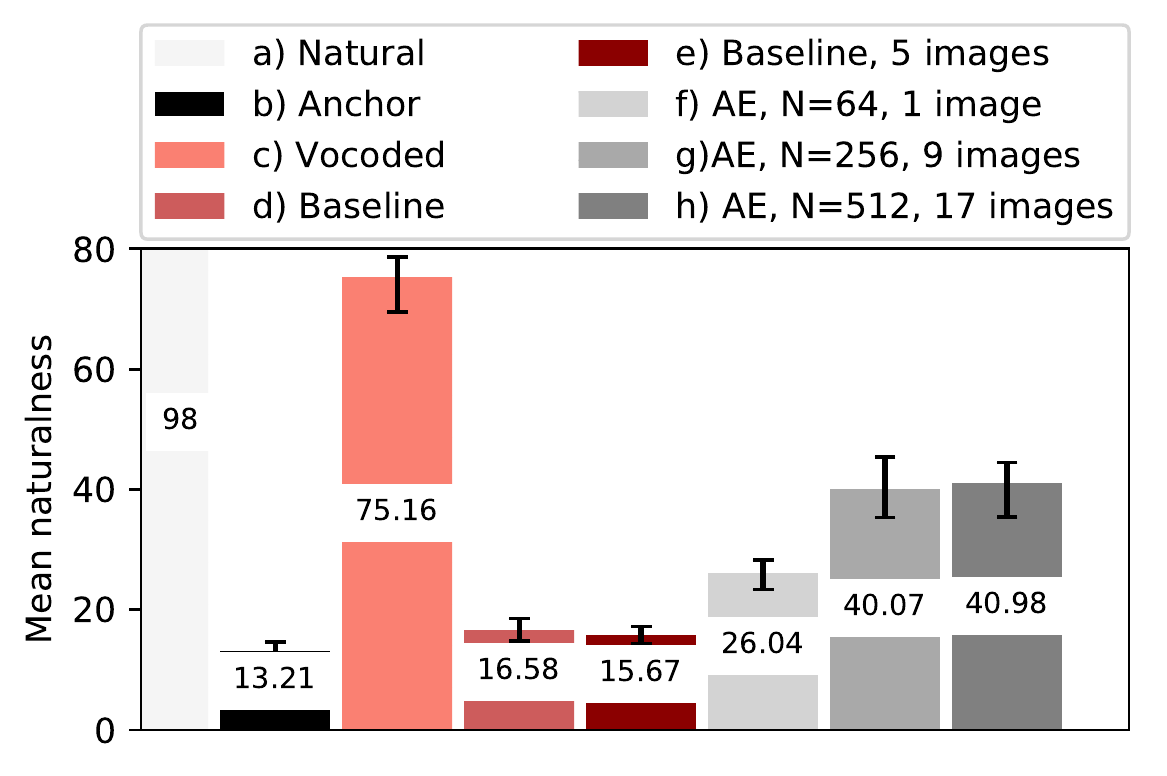}
\caption{Results of the listening test concerning naturalness. The
error bars show the 95\% confidence intervals.}
\label{results_mushra}
\end{figure}

Our listening test contained utterances synthesized from seven
variants of spectral estimates along with the reference recording.
Firstly, we used an {\it anchor} sentence, which was synthesized
from a distorted version of the original MGC-LSP features (i.e., in
analysis-synthesis with the vocoder, the lowest 6 values of the
MGC-LSP parameters were used from the original recording, while the
higher parameters were constant, resulting in a speech-like but
difficult-to-understand lower anchor). The {\it vocoded} reference
sentences were synthesized by applying impulse-noise excitation
using the original F0 and MGC-LSP values of the signals; these
utterances correspond to a form of ``glass ceiling'' for our DNN
models, and measure the loss of naturalness due to the speech
synthesis (i.e. vocoding) step. Next, we included both variants of
the baseline approach in the listening step, i.e. we used the 8\,192
pixels of the ultrasound images as features. In the first case, we
used only one image as input, while in the second one we
concatenated the pixels of five consecutive images. Lastly, we
included three autoencoder-based models in the listening test. The
first one was the simplest and smallest autoencoder-based model,
i.e. $N=64$ without any neighbouring vectors. As the other extreme
case, we tested the variation that had the most parameters, i.e.
$N=512$ using 8-8 neighbouring frames (17 frames overall). As the
last model tested, we chose the one that we found gave practically
optimal performance along with a (relatively) small number of
parameters: $N=256$ using 4-4 neighbouring feature vectors on both
sides.

Fig.~\ref{results_mushra} shows the average naturalness scores for
these tested approaches. In general, these values are in accordance
with the trends we found for the objective measurements. The
standard, pixel-by-pixel approach was the worst DNN-based technique
tested, and the subjects did not hear any improvement in the
naturalness of the synthesized samples when we used the neighbouring
frames as well to assist MGC-LSP prediction. (We would like to note,
though, that the synthesized sentences were understandable in each
case, even for the {\it anchor} approach.) Compared to the baseline
scores, using autoencoders for feature extraction brought
significant improvements: even the $N=64$ case without the help of
neighbouring frames led to an average naturalness score of 26.04\%.
The two further cases included in the listening test, i.e. $N=256$
with 9 frames and $N=512$ with 17 frames led to even more
natural-sounding synthesized utterances; our participants, however,
found no significant difference between these two configurations. Of
course, there is still room for improvement in the quality of the
resulting speech samples, as all the models tested produced clearly
lower quality utterances than the {\it vocoded} one.

\section{Conclusions}

We investigated the applicability of autoencoder neural networks in
ultrasound-based speech silent interfaces. In the proposed approach,
we used the activations of the bottleneck layer of the autoencoder
network as features, and we estimated the MGC-LSP parameters of the
speech synthesis step via a second deep network. According to our
experimental results, the proposed autoencoder-based process is a
more viable approach than the baseline one, which treats each pixel
as an independent feature: the estimations were more accurate in
every case, and the DNN model had fewer weights as well. Our
listening tests also demonstrated the benefit of using
autoencoder-based compression.

In our opinion, this improvement is mainly due to two factors.
Firstly, the autoencoder network automatically performs a de-noising
step on the input ultrasound image; as ultrasound videos are quite
noisy by nature, a de-noising step might help in the location of the
tongue and the lips, thus allowing more precise spectral parameter
estimation. The second advantage of our process is that the
autoencoder network also performs a compression of the original
image. Using the activations of the bottleneck layer significantly
reduced the size of our feature vector, which allowed us to estimate
the spectral speech synthesis parameters using more consecutive
images (i.e. a larger sliding window size) without relying on an
unrealistically huge feature vector.

We have several straightforward possibilities for continuing our
experiments. We could combine the autoencoder network with
convolutional neural networks, which will hopefully improve the
efficiency of the proposed procedure even more. An autoencoder-based
process can be also expected to tolerate slight changes in the
recording equipment position more than the baseline approach, where
we treat all pixels as independent features. Therefore, utilizing
the encoder part of the autoencoder network for feature extraction
might contribute to the development of more session-independent and
speaker-independent silent speech interface systems. We also plan to
perform these kinds of experiments in the near future.

%\begin{figure}[htbp]
%\centerline{\includegraphics{fig1.png}}
%\caption{Example of a figure caption.}
%\label{fig}
%\end{figure}

\section*{Acknowledgments}

L\'aszl\'o T\'oth was supported by the J\'anos Bolyai Research
Scholarship of the Hungarian Academy of Sciences and the UNKP-18-4
New Excellence Program of the Hungarian Ministry of Human
Capacities. Tam\'as Gr\'osz was supported by the National Research,
Development and Innovation Office of Hungary through the Artificial
Intelligence National Excellence Program (grant no.:
2018-1.2.1-NKP-2018-00008). We acknowledge the support of the
Ministry of Human Capacities, Hungary grant 20391-3/2018/FEKUSTRAT.
The authors were partially funded by the NKFIH FK 124584 grant and
by the MTA Lend\"ulet program. The Titan X graphics card used in
this research was donated by the Nvidia Corporation. We would also
like to thank the subjects who participated in the listening test.

%
%\section*{References}

\bibliographystyle{IEEEbib}
\bibliography{2019-ijcnn-ssi,ref_collection_csapot_nourl}

%\begin{thebibliography}{00}
%\bibitem{b1} G. Eason, B. Noble, and I. N. Sneddon, ``On certain integrals of Lipschitz-Hankel type involving products of Bessel functions,'' Phil. Trans. Roy. Soc. London, vol. A247, pp. 529--551, April 1955.
%\bibitem{b2} J. Clerk Maxwell, A Treatise on Electricity and Magnetism, 3rd ed., vol. 2. Oxford: Clarendon, 1892, pp.68--73.
%\bibitem{b3} I. S. Jacobs and C. P. Bean, ``Fine particles, thin films and exchange anisotropy,'' in Magnetism, vol. III, G. T. Rado and H. Suhl, Eds. New York: Academic, 1963, pp. 271--350.
%\bibitem{b4} K. Elissa, ``Title of paper if known,'' unpublished.
%\bibitem{b5} R. Nicole, ``Title of paper with only first word capitalized,'' J. Name Stand. Abbrev., in press.
%\bibitem{b6} Y. Yorozu, M. Hirano, K. Oka, and Y. Tagawa, ``Electron spectroscopy studies on magneto-optical media and plastic substrate interface,'' IEEE Transl. J. Magn. Japan, vol. 2, pp. 740--741, August 1987 [Digests 9th Annual Conf. Magnetics Japan, p. 301, 1982].
%\bibitem{b7} M. Young, The Technical Writer's Handbook. Mill Valley, CA: University Science, 1989.
%\end{thebibliography}
%\vspace{12pt}
%\color{red}
%IEEE conference templates contain guidance text for composing and formatting conference papers. Please ensure that all template text is removed from your conference paper prior to submission to the conference. Failure to remove the template text from your paper may result in your paper not being published.

\end{document}